# Graphene/SOI-based self-powered Schottky barrier photodiode array

A. Yanilmaz [1,2,3], M. Fidan[1], O. Unverdi[4] and C. Çelebi[1a)]

[1]Quantum Device Laboratory, Department of Physics, Izmir Institute of Technology, Izmir, 35430, Turkey

[2]Department of Photonics, Izmir Institute of Technology, Izmir, 35430, Turkey

[3]Ermaksan Optoelectronic R&D Center, Bursa, 16140, Turkey

[4]Faculty of Engineering, Department of Electrical and Electronic Engineering, Yasar University, Izmir, 35100, Turkey

We have fabricated 4-element Graphene/Silicon on Insulator (SOI) based Schottky barrier photodiode array (PDA) and investigated its optoelectronic device performance. In our device design, monolayer graphene is utilized as common electrode on lithographically defined linear array of n-type Si channels on SOI substrate. As revealed by wavelength resolved photocurrent spectroscopy measurements, each element in the PDA structure exhibited a maximum spectral responsivity of around 0.1 A/W$^{-1}$ under self-powered operational mode. Time-dependent photocurrent spectroscopy measurements showed excellent photocurrent reversibility of the device with ~1.36 μs and ~1.27 μs rise time and fall time, respectively. Each element in the array displayed an average specific detectivity of around $1.3 \times 10^{12}$ Jones and a substantially small noise equivalent power of ~0.14 pW/Hz$^{-1/2}$. The study presented here is expected to offer exciting opportunities in terms of high value-added graphene/Si based PDA device applications such as multi-wavelength light measurement, level metering, high-speed photometry, position/motion detection, and more.

______________________________

[a)] **Electronic mail:** cemcelebi@iyte.edu.tr

The photodiodes based on Graphene/Silicon (G/Si) heterojunction have attracted a great deal of attention in the last decade since they exhibit photo-responsivity similar to p-n or p-i-n type Si photodiodes[1,2]. It has been shown that a rectifying Schottky contact with an energy barrier level of around 0.5–0.8 eV is formed when graphene is laid on the surface bulk Si substrate[3,4]. Because of the fact that G/Si heterojunction operates as a Schottky barrier diode which is sensitive to light in the spectral range between 400–1100 nm due to the bandgap of Si. When graphene is employed as an electrode on Si, it does not only act as optically transparent conductive layer, but it functions also as a photon absorbing active material similar to metal silicide electrodes used in conventional metal/Si Schottky barrier photodiodes[5,6]. Under light illumination, although a large amount of photons is converted into photo-generated charge carriers in Si, the optical absorbance in graphene (~2.3 %) contributes to light detection as well through internal photoemission over the Schottky barrier. The photo-generated electrons, with high enough energies to overcome the Schottky barrier, are accelerated towards the bulk of Si due to the built-in electric field at the interface of G/Si. As the electrons pass through the depletion region of Si, they undergo energy loss processes that prevent them from passing back into the graphene over layer. This results in an effective charge separation and hence a measurable photocurrent and/or photovoltage in such G/Si photodiodes even under zero-bias conditions.

A variety of different design strategies have been employed to fabricate single element G/Si based Schottky barrier photodiodes. These include the transfer of monolayer graphene either on oxide-free or on nanotip patterned surface of bulk Si substrates[7,8]. There are several other approaches relying on the fabrication of these type of photodiodes using silicon-on-insulator (SOI) technology. SOI structure, which is composed of a buried $SiO_2$ (BOX) sandwiched between a thin Si layer and a thick Si substrate, provides great opportunities for producing G/Si based photodetectors with improved device performance. For example, a bottom-gated SOI transistor with isolated patterned graphene layers on top of a single channel Si has been utilized as a single pixel photodetector to detect light in the visible to the near-infrared range[9]. In a subsequent work, it has been shown that SOI based single pixel G/Si Schottky photodiodes exhibit a maximum spectral responsivity of around 0.26 AW$^{-1}$ at 635 nm peak wavelength and a response time substantially smaller than a microsecond compared to their counterparts fabricated on bulk Si substrates[10].

In this letter, we demonstrate a multi-channel G/Si Schottky barrier photodiode array (PDA) can be fabricated on SOI substrates using standard microfabrication techniques applied in CMOS technology. In our device design, we used the advantage of the BOX layer in SOI which acts as a well-defined etch-stop and provides an excellent electrical isolation in between laterally aligned neighboring photoactive G/Si elements in the array. In the fabrication process, single layer graphene is utilized as common electrode on a linear array of multiple n-type Si channels which were lithographically exposed on a single SOI substrate. Current-voltage (I–V) and wavelength resolved photocurrent spectroscopy measurements showed that each G/Si element in the PDA operates in self-powered mode and responds to incident light independently of each other. The optoelectronic device parameters including spectral responsivity, specific detectivity, noise equivalent power and response speed of the G/Si PDA sample were systematically investigated and reported in the letter.

Fig. 1(a) shows a schematic illustration of the SOI based 4-element G/Si Schottky PDA fabricated within the scope of this work. The details of graphene growth, characteristic dimensions of Si elements on SOI and device fabrication steps can be found in the supplementary material. For the experiments 10 mm × 10 mm sized, 10 μm thick n-doped photo-active silicon (Si (100)) SOI substrates (specification $\rho$ = 1–5 $\Omega$.cm, nominal doping level $N_d \approx 2 \times 10^{15}$ cm$^{-3}$) were used. The device structures were prepared by using photolithography technique. Dry etching method (Reactive Ion Etching (RIE), Sentech Ins.) was used to reach the oxide layer (BOX) and to obtain an array of n-Si channels on SOI substrates. Following the fabrication of Si array, the windows for metal contact pads were defined by an additional lithography step. After Cr (5 nm)/Au (80 nm) metals were evaporated both on the n-Si side and on the SiO$_2$ side of the SOI substrates with a thermal evaporation system and then a lift-off process was applied to obtain linear array device structure as schematically depicted in Fig.1(a). Chemical Vapor Deposition (CVD) grown monolayer graphene with a surface coverage of higher than 95 % was transferred on the arrayed SOI substrate by using the same graphene transfer method in Ref [11]. The graphene layer in the device design was employed as hole collecting common electrode and acted as the active region when interfaced with the arrayed n-Si elements on SOI substrate.

After the transfer process, the presence of graphene on Si and SiO$_2$ regions of the SOI substrate were determined by single point Raman spectroscopy measurements taken under a laser with 532 nm excitation wavelength. As shown in Fig. 1(b), graphene related D, G, and single Lorentzian shaped 2D peaks were identified in all the obtained Raman spectra. Strong G peak and weak D peak indicate good graphitic quality, and the ratio of 2D peak intensity with G peak intensity ($I_{2D}/I_G > 2$) confirms that graphene is monolayer[12] on arrayed SOI substrate.

Following the Raman analysis, I–V measurements of each G/n-Si element on the SOI substrate were conducted one by one under dark conditions and the obtained results are plotted in Fig. 1(c). All the I–V curves exhibit strong rectifying character of a typical Schottky barrier diode but with slightly different current levels varying in the forward bias region (Fig.1(c)). For a detailed comparison, the I–V data were plotted in the semi-logarithmic scale as shown in Fig. 1(d). From the ln(I)–V plots, the dark current ($I_d$) of the G/n-Si elements were extracted as ~0.5 nA in average. In the forward-bias range, although the current increases linearly at very small voltages in accordance with the well-known thermionic-emission model, the deviation from linearity observed at relatively high voltages (e.g., $V_b > 0.1$ V) is due to the series resistance contributions from the underlying n-Si element. The slight difference seen at the reverse bias saturation currents suggests only a small variation in the rectification strength of the G/n-Si heterojunction. Using the method developed by Cheung et al.[13], the average Schottky barrier height ($\Phi_B$) and the ideality factor ($\eta$) of the PD elements were extracted from the linear forward-bias region of the ln(I)–V plot as 0.78 eV and 1.48, respectively. These two diode parameters are consistent with those of G/Si based Schottky barrier photodiodes fabricated on thick and bulk n-Si substrates[14,15].

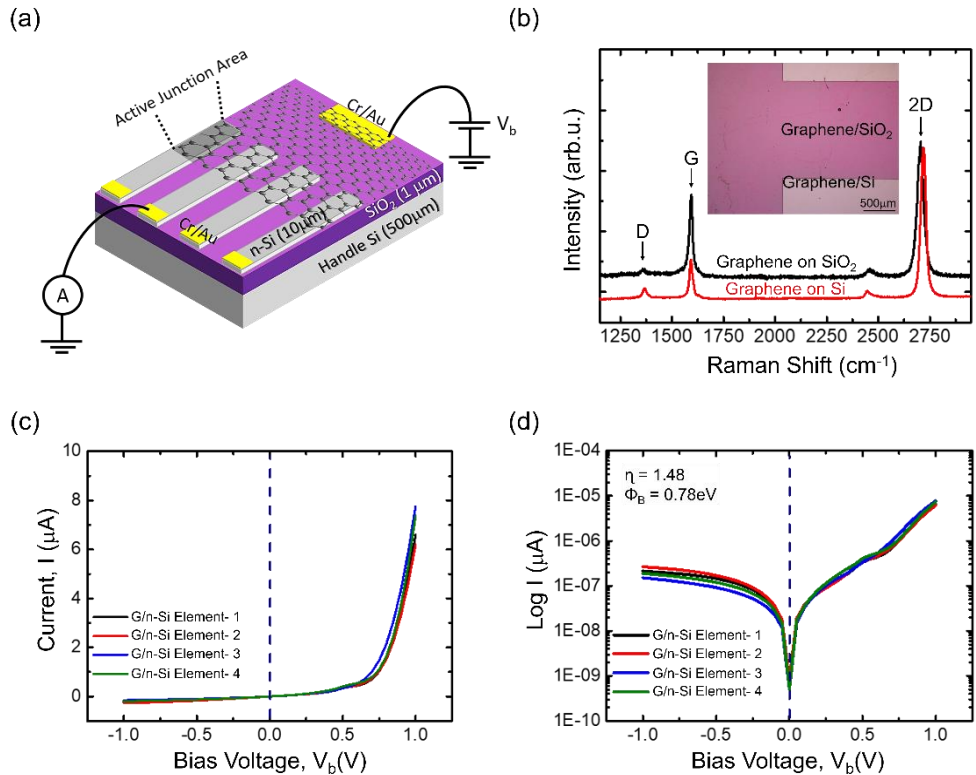

FIG. 1. (a) The schematic of the fabricated G/n-Si PDA device. (b) Raman spectrum of transferred graphene on Si and $SiO_2$ sides of SOI substrate (Inset shows the optical micrograph of the region selected for acquiring a Raman spectrum of the graphene layer). (c) The I-V curve of the device in dark and (d) the ln(I)-V plot which was used to extract the ideality factor and Schottky barrier height of each element.

The photoresponse of each individual G/n-Si element in the PDA was characterized separately under illumination of light with 660 nm wavelength. For the experiments an LED source was coupled to a fiber optic cable with 600 μm core diameter to maintain local illumination on each G/n-Si element and on SiO$_2$ regions between them as depicted in Figs. 2(a) and (b). To avoid possible optical cross-talk between neighboring n-Si elements, the distance between the sample and the tip of the fiber optic cable was kept at ~1 mm to ensure a well-defined spot size and to minimize possible back reflections that may respectively arise from the illuminated Si surface and the metallic tip of the fiber optic probe used in the experiments. As seen in the ln(I)–V plots (Fig. 2(c)), all the G/n-Si elements displayed a clear photoresponse with measurable photocurrent ($I_{pc}$) under light illumination even at zero-bias ($V_b$ = 0 V). The shift of the minimum current seen at forward bias range corresponds to the open circuit voltage ($V_{oc}$) and is typical for self-powered G/n-Si photodiodes operating in the self-powered mode[16]. It is known that, when G/n-Si heterojunction is subject to light illumination, the incident photons pass through the optically transparent graphene electrode are absorbed by the Si substrate underneath. As a result of photo excitation, electron–hole pairs are created at the depletion region. The photo-generated charge carriers are separated due to an effective built-in electric field at the interface between graphene layer and n-Si. Optically excited charge carriers in graphene gain sufficient energy to overcome the Schottky barrier formed at the G/n-Si interface and lead to a photocurrent[17] even at zero-bias. From the I–V plot shown in Fig. 2(c), zero-bias $I_{pc}$ of G/n-Si elements were determined to be varying in a range between 1.6 and 3.1 μA. In the case when the light source is brought on the $SiO_2$ regions located in between two neighboring active elements (Fig. 2(b)), the zero-bias currents of G/n-Si elements were measured as ∼ 3.1–9.4 ×$10^{-9}$ A. Compared to the corresponding $I_d$ values, such a slight increase in the measured currents is due to a trace amount of light which was randomly reflected back from the tip of the metallic casing of the fiber optic probe onto the surface of photoactive G/n-Si elements. When the effects of reflected light on the measured current are ignored, it is possible to state that there is almost no cross-talk between neighboring G/n-Si elements in the array.

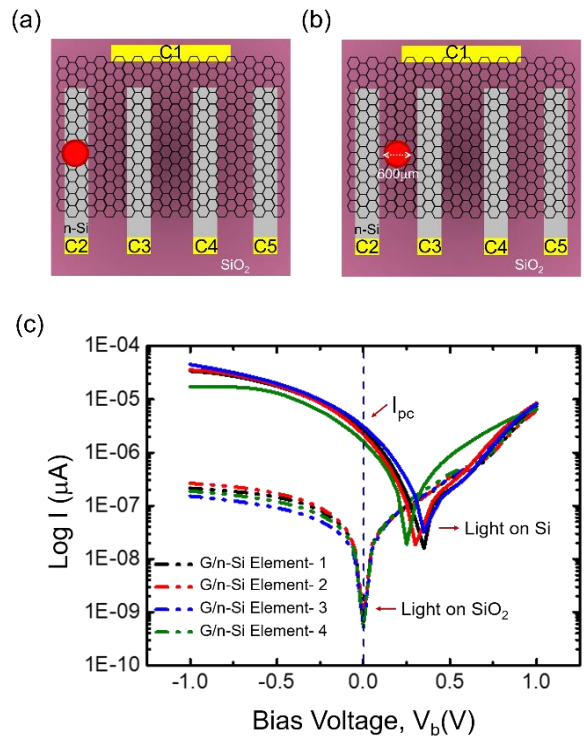

FIG. 2. Optical cross-talk measurement results of G/n-Si Schottky PDAs; (a) local illumination on each G/n-Si element, (b) illumination on SiO2 regions located in between two neighboring active elements and (c) ln(I)–V measurements acquired on G/n-Si elements and on SiO2 regions between them. (C1; common graphene contact, and C2, C3, C4 and C5 represent the contacts on n-Si).

The spectral responsivity (R) is one of the most important parameters for the light sensing capability of a photodiode and can be written as[18]

$$R = \frac{I_{pc} - I_d}{P(\lambda)} \qquad (1)$$

where P is the optical power of incident light at a certain wavelength (λ). The spectral responsivity of each element in the PDA was measured at $V_b$ = 0 V as a function of λ of the incident light varied in the spectral range between 400 and 1050 nm. The obtained results were compared with those of a typical G/n-Si based reference PD fabricated on a 500 μm thick bulk n-Si having the same doping concentration as n-Si layer on the SOI substrate. As seen in Fig. 3(a), the maximum spectral responsivity ($R_{max}$) of the reference PD and G/n-Si element were determined as ∼0.7 AW$^{-1}$ and ∼0.1 AW$^{-1}$, respectively. The difference in the maximum spectral responsivities is due to the active junction area of the reference PD (~20 mm$^2$) which is larger than that of each G/n-Si element (~3 mm$^2$) in the array. It has been shown that larger active junction area leads to a wider depletion region which promotes the effective separation/collection of the photo-generated charge carriers at the depletion region. As a consequence of enhanced charge separation efficiency, $I_{pc}$ and R increase proportional with the size of the active junction area[19].

For a detailed comparison, normalized spectral responsivities of the reference PD and a typical G/n-Si element were plotted in Fig. 3(b). Different from that of the reference PD, the spectral responsivity of G/n-Si element exhibits two maxima located at around 660 nm and 780 nm wavelengths and decays earlier for the wavelengths above 780 nm. The observed difference in the two distinct spectral responsivity characteristics can be understood in terms of the penetration depth of light and reduced absorption coefficient of Si layer on SOI. The photoresponse contribution is only provided by the absorption of incident light in active thin Si layer on SOI when compared with bulk Si. The BOX layer and Si handle make ineffective the remaining light power for photoresponse gain in SOI structure[10]. Accordingly, light-trapping concepts that prolong the light path in thick and bulk Si substrates should be considered. Although only the drift currents contribute in shorter wavelengths, diffusion currents become dominant in longer wavelength regime where the light penetrates deeper into the substrate. For G/Si Schottky photodiodes on bulk Si, in which the depletion region is wider compared to that of G/n-Si element out of thin Si, the spectral responsivity is shifted towards longer wavelengths due to increased amount of the diffusion currents[7]. Because of the fact that the spectral responsivity of G/n-Si elements in the array appear to decline earlier than that of G/Si PD fabricated out of thick and bulk n-Si substrates[20]. As also reported for SOI based single pixel G/Si Schottky PD[10], the photo-active thin Si forms an optical microcavity on the layered structure of the SOI substrate and causes an oscillating spectral responsivity as displayed in Fig. 3(b). Such an oscillatory behavior arise from constructive and destructive interference effects due to multiple reflections occurred between different interfaces in the SOI structure[10].

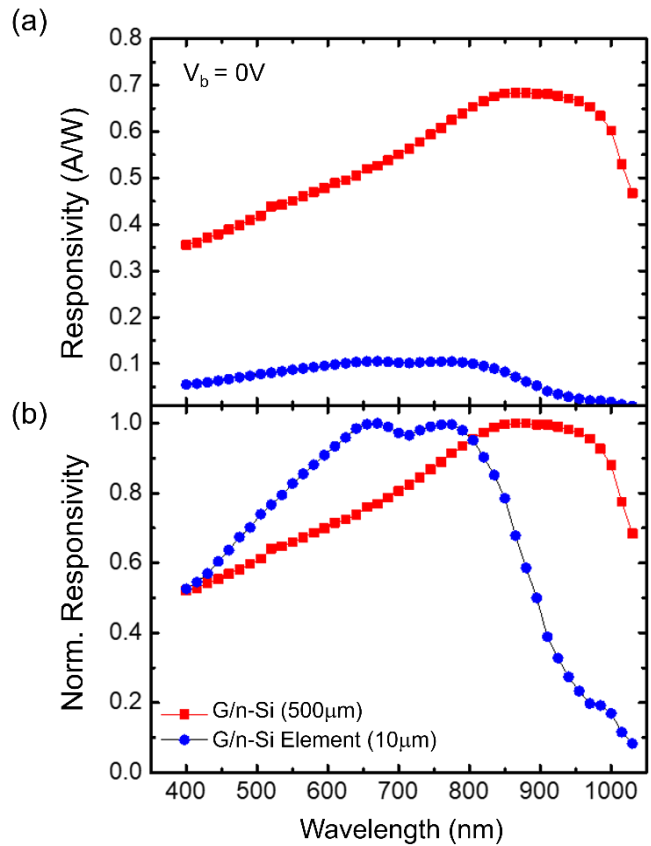

FIG. 3. (a) Comparison of the spectral responsivities of a typical G/n-Si element in the PDA device on SOI substrate (blue) and a reference G/Si Schottky PD fabricated on a 500 μm thick bulk Si substrate (red) under zero-bias voltage. (b) Comparison of the normalized spectral responsivities of these two devices.

Taking into account the maximum spectral responsivity ($R_{max}$) read at 660 nm wavelength, we also calculated the specific detectivity (D*) and noise equivalent power (NEP) parameters of each active element in the array. Here, D* is defined as the weakest level of light detected by a photodiode with a junction area (A) of 1 cm² and is determined by[20,21]

$$D^* = \frac{A^{1/2} R}{\sqrt{2eI_d}} \quad (2)$$

and NEP is the incident power required to obtain a signal-to-noise ratio of 1 at a bandwidth of 1 Hz and is calculated by[19,22]

$$NEP = \frac{A^{1/2}}{D^*} \quad (3)$$

Considering $R_{max}$ = 0.1 AW$^{-1}$ and A = 3 mm², the average D* and NEP of the G/n-Si elements were calculated as ~1.3 × 10$^{12}$ Jones and ~0.14 pW/Hz$^{-1/2}$, respectively. These two photodiode parameters are in good agreement with those of both single pixel G/Si PD on SOI and G/Si PD on bulk Si substrates[10,19].

In order to determine the response speed and 3-dB bandwidth ($B_w$) of the active elements in the PDA, we conducted time-resolved photocurrent spectroscopy measurements using the experimental set-up illustrated in Fig. 4(a). The measurements revealed that all the elements have excellent photocurrent on/off reversibility as seen in Fig. 4(b). Rise time ($t_r$) and decay time ($t_d$) of each individual element in the array were determined from single pulse response measurements taken under 660 nm wavelength light pulsed with 1 kHz frequency. Here $t_r$ is defined as the range that the photocurrent rises from 10 to 90 % of its maximum and $t_d$ is defined similarly. A typical example for single pulse response measurement on an individual G/n-Si element is shown in Fig. 4(c). Considering the measurements taken on other individual G/n-Si elements in the PDA, the average $t_r$ and $t_d$ were determined as ~1.36 μs and ~1.27 μs, respectively. Using the relation $B_w = 0.35/t_r$, the average 3-dB $B_w$ of the G/n-Si elements in the array was calculated as ~257 kHz. For convenience, all the obtained performance characteristics of each element in the PDA device structure are listed in Table 1. The comparison of key parameters such as the preferable R, D*, and NEP, low noise, rapid response time of G/n-Si PDA obtained (under zero bias voltage) in this work with previous reported G/Si PDs can be found in Table S1 of the supplementary materials.

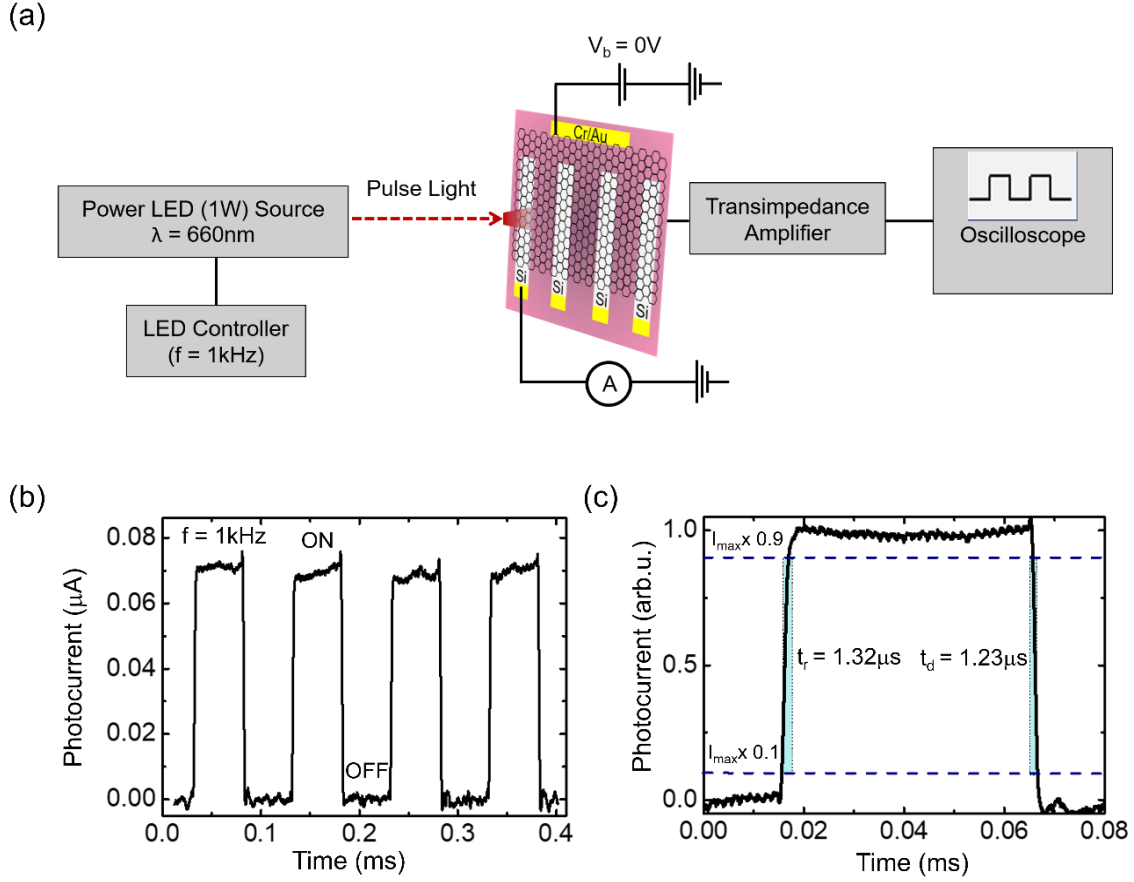

FIG. 4. (a) Schematic diagram of time-resolved photocurrent spectroscopy measurement setup. (b) Time-resolved photocurrent spectrum of an individual G/n-Si element under 660 nm wavelength light with 1 kHz switching frequency at zero-bias voltage. (c) One cycle time-resolved photocurrent spectrum of an individual G/n-Si element in the PDA device.

TABLE 1. Summary of the performances of the 4-Element G/n-Si Schottky PDAs under 660 nm wavelength light at 0 V bias voltage (Junction area 3mm$^2$).

| Element ID | $I_{dark}$ (nA) | $R_{max}$ (A/W) | $D^*$ ($10^{12}$) (Jones) | NEP (pW/Hz$^{-1/2}$) | $t_r$ (μs) | $t_d$ (μs) | 3-dB $B_w$ (kHz) |
|---|---|---|---|---|---|---|---|
| G/n-Si Element -1 | 0.6 | 0.11 | 1.38 | 0.125 | 1.40 | 1.28 | 250 |
| G/n-Si Element -2 | 0.8 | 0.10 | 1.29 | 0.161 | 1.38 | 1.31 | 253 |
| G/n-Si Element -3 | 0.5 | 0.10 | 1.26 | 0.127 | 1.32 | 1.23 | 265 |
| G/n-Si Element -4 | 0.3 | 0.09 | 1.25 | 0.138 | 1.33 | 1.21 | 263 |

In conclusion, we have demonstrated G/n-Si based multi-channel Schottky barrier linear PDA on a conventional SOI substrate and investigated its optoelectronic characteristics. In the PDA device structure, monolayer graphene is utilized as common electrode on 4-element n-Si array fabricated with two-step photolithography process on a single SOI structure. The I–V measurements taken under dark ambient revealed a rectifying Schottky contact between graphene electrode and each n-Si element in the array. Each of the individual element in the array exhibited a clear photo-response even under (zero-bias) self-powered operational mode similar to that observed for graphene/SOI based single pixel Schottky barrier photodiode which was previously reported in literature. We showed that multiple G/n-Si photodiodes, operating independently from each other on a single SOI substrate, can be produced by employing graphene as common electrode. This study offers exciting opportunities for the realization of high-value added technological applications based on motion and position detection, imaging and spectrophotometry in which graphene and SOI technology can be used together.

See the supplementary material for further experimental details, device fabrication process along with corresponding figures and comparison of G/n-Si Schottky barrier PDA with previously reported single element G/n-Si photodiodes fabricated on bulk Si and on SOI substrates.

## AUTHOR DECLARATIONS

### Conflict of Interest

The authors declare no conflict of interest.

### DATA AVAILABILITY

The data that support the findings of this study are available from the corresponding authors upon reasonable request.

### ACKNOWLEDGEMENTS

The authors would like to thank the researchers in Center for Materials Research of İzmir Institute of Technology (İYTE MAM) and Ermaksan Optoelectronic R&D Center in Turkey for their support in device fabrication processes. This work was supported within the scope of the scientific research project which was accepted by the Project Evaluation Commission of Yasar University under the project number and title of "BAP113_Grafen Elektrotlu SOI Tabanlı Doğrusal Fotodedektör Dizisi Geliştirilmesi".

# Supplementary Material

Graphene/SOI based self-powered Schottky barrier photodiode array

A. Yanilmaz [1,2,3], M. Fidan[1], Ozhan Unverdi[4] and C. Çelebi[1*]

[1]*Quantum Device Laboratory, Department of Physics, Izmir Institute of Technology, Izmir, 35430, Turkey*

[2]*Department of Photonics, Izmir Institute of Technology, Izmir, 35430, Turkey*

[3]*Ermaksan Optoelectronic R&D Center, Bursa, 16140, Turkey*

[4]*Faculty of Engineering, Department of Electrical and Electronic Engineering, Yasar University, Izmir, 35100, Turkey*

[*]E-mail: cemcelebi@iyte.edu.tr

**Supplementary Information**

1. Graphene Growth

2. Fabrication of 4-Element n-Si PDAs

3- Optoelectronic Characterization

4- Comparison of the properties of our device with recently developed graphene-based photodetectors.

**1- Graphene Growth**

In this work, self-limiting growth of monolayer graphene with a surface coverage of higher than 95% was grown on a high purity, on 25 μm thick unpolished copper foil (99.8 purity, Alfa Aesar) by Atmospheric Pressure Chemical Vapor Deposition (APCVD) method. As for the first step, the Cu foil was heated up to 1000°C under a $H_2$ (20 sccm) + Ar (1000 sccm) gas mixture with a temperature ramp rate of 30 °C min$^{-1}$. Then the foil was annealed under the same temperature and flow rates for one hour. After the annealing process, $CH_4$ (10 sccm) was introduced into the tube furnace for two minutes in order to facilitate the graphene growth. Finally, the sample was left for rapid cooling from growth temperature to room temperature under gas flows of $H_2$ (20 sccm) and Ar (1000 sccm). Temperature–Time diagrams of the graphene growth was represented in **Figure S1**. In this study, large area monolayer graphene was grown on ~ 60 mm$^2$ Cu foil.

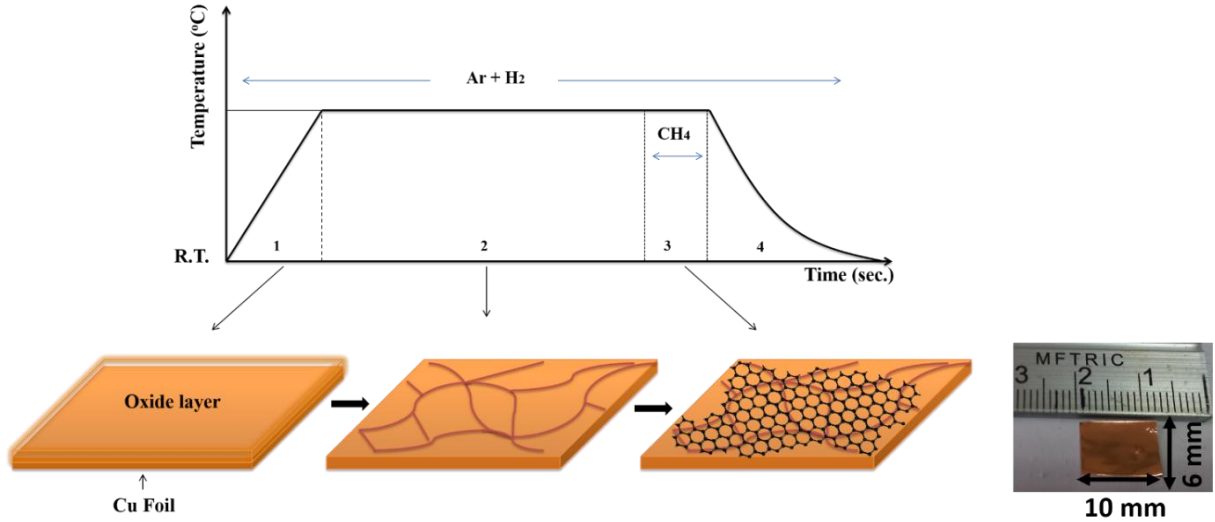

**Figure S1**. Temperature–time diagrams of the graphene growth. Heating (1), annealing (2), growth (3) and cooling (4) stages of graphene growth process on Cu foil, respectively.

## 2- Fabrication of 4-Element G/n-Si PDA on SOI

The 4-Element structures were prepared by using maskless lithography system (Heidelberg µMLA). Following the fabrication of Si arrays, the metal contacts were fabricated by an additional lithography step. After the Cr (5 nm)/Au (80 nm) metal contact pads were evaporated both on the n-Si side and on the $SiO_2$ covered the side of SOI substrates with the thermal evaporation system, the liftoff process was done to obtain our devices (**Figure S2(a)**). After graphene growth, as the supporting layer Microposit S1318 Photoresist (PR) was drop casted on the Gr/Cu and the stack was annealed at 70°C overnight in an oven. The Cu foil at the bottom was fully etched using Iron Chloride ($FeCl_3$) solution to get suspended PR/Gr bilayer. After etching of possible $FeCl_3$ residues in $H_2O$: HCl (3:1) solution, the PR/Gr became ready for transfer process (**Figure S2(b)**). Monolayer CVD grown graphene was employed as *common* Schottky electrode and acted as the active region when interfaced with arrayed n-Si substrate. The active area of the junction is determined by with the area of 3 $mm^2$ where silicon and graphene meet on arrayed Si (**Figure S2(c)**). Each element has a Si dimensions with length and width of 5mm and 1mm, respectively. The length between elements is kept constant as 1.5 mm (**Figure 2(d)**).

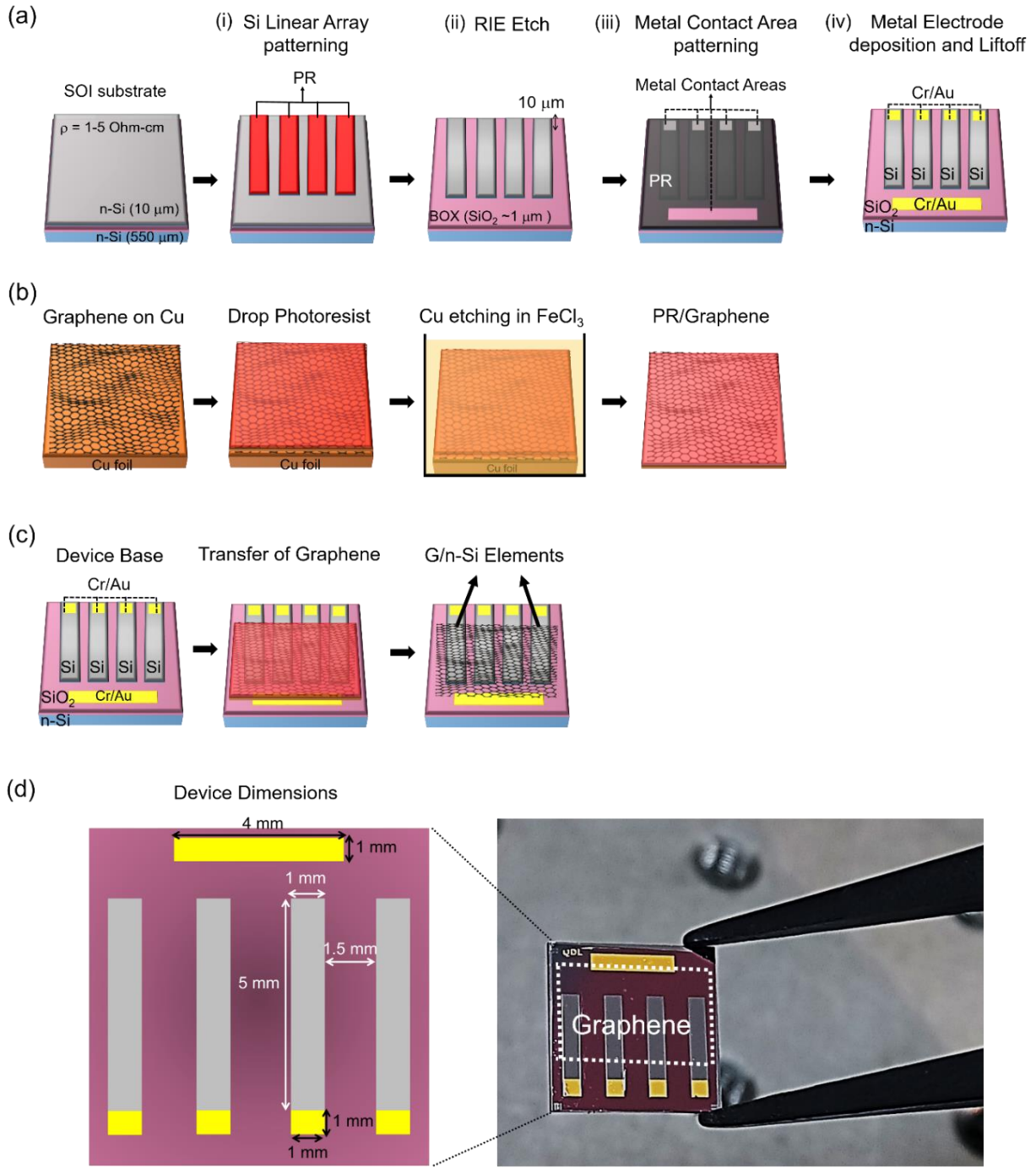

**Figure S2**.(a) Fabrication steps of 4-Element n-Si PDAs, (b) schematic illustration of CVD graphene growth on Cu foil and its transfer preparation onto arrayed Si substrate, (c) 4-Element G/n-Si Schottky PDAs device base, followed by transferring of graphene, (d) device dimensions and device picture. (The active junction area of the elements is 3 mm$^2$).

## 3- Optoelectronic Characterization

For two-terminal I–V measurements, 80 μm thick Cu wires were bonded both on the Cu plated paths of a PCB card and Cr/Au contact pads by means of a lab-built wire bonder. In order to perform optoelectronic characterizations of 4-Element G/n-Si Schottky PDAs, tungsten-halogen lamp (Osram, 275 W) was used to generate light and specific wavelengths were separated with the help of a monochromator (Newport, Oriel Cornerstone) including internal shutter. A spectrophotometer (Oceans Optics) was used to calibrate the full width of half maximum of light by changing the slit of spectrophotometer and power output of commercial Si Photodiode FDS10X10 (Thorlabs) was obtained to define the incoming power of light on device area. Then responsivity vs wavelength (resolution 15 nm) measurements were employed by using Keithley 2400 source-meter and Keithley 6485 picoammeter. In addition to the systems mentioned above, the time-resolved photocurrent measurements of the devices of all devices was evaluated for deep red wavelengths ranging employing 660 nm 1W Power LED and function generator as LED driver with pulse modulation (Uni-t utg9005c). The photoresponse measurements of devices were acquired under 660 nm wavelength light pulsed with 1 kHz frequency and optical crosstalk measurements were taken using a 600 μm diameter fiber optic cable. The irradiation wavelength is specifically selected to be 660 nm since it corresponds to the maximum spectral response of our fabricated G/n-Si elements on 10μm thick SOI substrates ($R_{max}$ = ~0.1 AW$^{-1}$).

## 4- Comparison of the properties of our device with recently developed graphene-based photodetectors.

**Table S1.** Key parameters comparison of 4-Element G/n-Si Schottky PDAs on SOI substrate obtained in this work with previous reported graphene/Si photodetectors.

| Types of Devices | Junction Area | Responsivity | Detectivity | Rise/Fall Time | refs |
|---|---|---|---|---|---|
| Graphene/n-Si | 4 mm$^2$ | 0.16 A/W ($V_b$ = 0 V; λ = 905 nm) | 0.37×10$^{13}$ Jones | 6.4μs /9.6μs | 1 |
| 3D-graphene/SOI | - | 27.4 A/W ($V_b$ = -0.5 V; λ = 1550 nm) | 1.37×10$^{11}$ Jones | 212μs /242μs | 2 |
| Graphene/silicon-on insulator in conductor mode | - | 10$^{17}$ A/W ($V_b$ = -30 V; λ = 532 nm) | 1.46×10$^{13}$ Jones | 90μs/- | 3 |
| Graphene–silicon-on insulator | 16 mm$^2$ | 0.26 A/W ($V_b$ = -2 V; λ = 635 nm) | 7.83×10$^{10}$ Jones | ~10ns/20-70ns | 4 |
| 4-Element G/n-Si Schottky PDA | 3 mm$^2$ | 0.10 A/W ($V_b$ = 0 V; λ = 660 nm) | 1.3x10$^{12}$ Jones | ~1.36μs /~1.27 μs | Our work |